\RequirePackage{fix-cm}
\documentclass[smallextended]{svjour3}       
\smartqed  
\usepackage{graphicx,amssymb,amsfonts,mathrsfs,latexsym,amsmath,amssymb}
\usepackage{color}
\journalname{AAECC}
\begin{document}

\title{Cyclic codes over $\mathbb{F}_{2^m}[u]/\langle u^k\rangle$ of oddly even length
} \subtitle{}

\titlerunning{Cyclic codes over $\mathbb{F}_{2^m}[u]/\langle u^k\rangle$ of oddly even length}

\author{Yonglin Cao$^1$ $\cdot$ Yuan Cao$^2$ $\cdot$ Fang-Wei Fu$^3$ 
}


\institute{Yonglin Cao \at
             \email{ylcao@sdut.edu.cn}\\
           Yuan Cao\at
             \email{yuan$_{-}$cao@hnu.edu.cn}\\
           Fang-Wei Fu\at
             \email{fwfu@nankai.edu.cn} \\
          $^1$  School of Sciences, Shandong University of Technology, Zibo, Shandong 255091, China \\
          $^2$  College of Mathematics and Econometrics, Hunan University, Changsha 410082, China\\
          $^3$  Chern Institute of Mathematics and LPMC, Nankai University, Tianjin 300071, China
          }

\date{Received: date / Accepted: date}

\maketitle

\begin{abstract}
Let $\mathbb{F}_{2^m}$ be a finite field of characteristic $2$ and $R=\mathbb{F}_{2^m}[u]/\langle u^k\rangle=\mathbb{F}_{2^m}
+u\mathbb{F}_{2^m}+\ldots+u^{k-1}\mathbb{F}_{2^m}$ ($u^k=0$) where $k\in \mathbb{Z}^{+}$ satisfies $k\geq 2$.
For any odd positive integer $n$, it is known that cyclic codes over $R$ of length $2n$ are identified with ideals of the ring
$R[x]/\langle x^{2n}-1\rangle$. In this paper, an explicit representation for each cyclic code over $R$ of length $2n$ is provided and a formula to count the number of codewords in each code is given. Then a formula to calculate the number of cyclic codes over $R$ of length $2n$
is obtained. Moreover, the dual code of each cyclic code and self-dual cyclic codes over $R$ of length $2n$ are investigated.

\keywords{Cyclic code \and Finite chain ring \and Non-principal ideal ring \and Dual code \and
Self-dual code
\vskip 3mm \noindent
{\bf Mathematics Subject Classification (2000)} 94B05 \and 94B15 \and 11T71}
\end{abstract}

\section{Introduction and preliminaries}
\label{intro}
Throughout this paper, let $\mathbb{F}_{p^m}$ be a finite field of characteristic $p$ and
cardinality $p^m$, and denote $R=\mathbb{F}_{p^m}[u]/\langle u^k\rangle=\mathbb{F}_{p^m}
+u\mathbb{F}_{p^m}+\ldots+u^{k-1}\mathbb{F}_{p^m}$ ($u^k=0$), where $p$ is a prime and $k\in \mathbb{Z}^{+}$ satisfying $k\geq 2$.
It is well-known that $R$ is a finite chain ring. For any positive integer $N$,
let $R[x]/\langle x^N-1\rangle$ be the residue class ring of the polynomial ring $R[x]$ modulo its ideal $\langle x^N-1\rangle$ generated by $x^N-1$.
Then every element of $R[x]/\langle x^N-1\rangle$ can be uniquely expressed as
$a(x)+\langle x^N-1\rangle$ where $a(x)=\sum_{j=0}^{N-1}a_jx^j\in R[x]$ with $a_0,a_1,\ldots,a_{N-1}\in R$.
As usual, we will identify $a(x)+\langle x^N-1\rangle$ with $a(x)$ for simplicity. Hence
\begin{center}
$R[x]/\langle x^N-1\rangle=\{\sum_{j=0}^{N-1}a_jx^j\mid a_0,a_1,\ldots,a_{N-1}\in R\}$
\end{center}
which is a ring with the operations
defined by the usual polynomial operations modulo $x^N-1$.
It is well-known that cyclic codes over
$R$ of length $N$ are identified with ideals of the ring $R[x]/\langle x^N-1\rangle$, under the
identification map $\theta: R^N \rightarrow R[x]/\langle x^N-1\rangle$ via $\theta: (a_0,a_1,\ldots,a_{N-1})\mapsto
\sum_{j=0}^{N-1}a_jx^j$ ($\forall a_0,a_1,\ldots,a_{N-1}\in R$).

\par
   There were many literatures on cyclic codes of length $N$ over finite chain rings $R=\mathbb{F}_{p^m}[u]/\langle u^k\rangle$ for various finite fields $\mathbb{F}_{p^m}$
and positive integers $k$ and $N$. For example: When $k=2$, Bonnecaze and Udaya [3] investigated cyclic codes and self-dual codes over $\mathbb{F}_2+u\mathbb{F}_2$ for odd length $N$. Norton [7] discussed cyclic codes of length $N$ over an arbitrary finite chain ring $R$ systematically when $N$ is not divisible by the characteristic of the residue field $\overline{R}$.
    Dinh [4] studied cyclic codes and constacyclic codes over
Galois extension rings of $\mathbb{F}_2+u\mathbb{F}_2$ of length $N=2^s$, and then investigated cyclic codes and constacyclic codes over
$\mathbb{F}_{p^m}+u \mathbb{F}_{p^m}$  of length $N=p^s$ in [5].

\par
   When $k\geq 3$, for the case of $p=2$ and $m=1$ Abualrub and Siap [1] studied cyclic codes over the ring $\mathbb{Z}_2+u\mathbb{Z}_2$ and $\mathbb{Z}_2+u\mathbb{Z}_2+u^2\mathbb{Z}_2$ for arbitrary length $N$, then Al-Ashker and Hamoudeh [2] extended some of the results in [1],
and studied cyclic codes of an arbitrary length over
the ring $Z_2+uZ_2+u^2Z_2+\ldots+u^{k-1}Z_2$ with $u^k=0$. The rank and minimal spanning set of this family of codes were studied and two open problems were
asked: \textsf{The study of cyclic codes of an arbitrary length over $Z_p+uZ_p+u^2Z_p+\ldots+u^{k-1}Z_p$,
where $p$ is a prime integer, $u^k=0$, and the study of dual and self-dual codes and their properties over these rings}.
Recently, for the case of $m=1$ Singh et al. [8] studied cyclic
code over the ring $\mathbb{Z}_p[u]/\langle u^k\rangle=Z_p+uZ_p+u^2Z_p+\ldots+u^{k-1}Z_p$ for any prime integer $p$ and positive integer $N$. A set of
generators, the rank and the Hamming distance of these codes were investigated. But for arbitrary prime $p$ and
positive integers $m, N$ and $k\geq 3$, the following questions for cyclic codes over $R$ have not been
solved completely to the best of our knowledge:

\par
  $\bullet$ For each cyclic code ${\cal C}$ over $R$ of length $N$, give a unique representation for
${\cal C}$ and provide a clear formula to count the number of codewords in ${\cal C}$. Then obtain a formula to count the number of all such cyclic codes.

\par
  $\bullet$ Using the representation of any cyclic code ${\cal C}$, give the representation
of the dual code of ${\cal C}$ precisely and determine the self-duality of ${\cal C}$.

\par
   From now on, let $R=\mathbb{F}_{2^m}[u]/\langle u^k\rangle=\mathbb{F}_{2^m}
+u\mathbb{F}_{2^m}+\ldots+u^{k-1}\mathbb{F}_{2^m}$ ($u^k=0$) where $k\geq 2$, and $n$ be an odd positive integer. We focus our attention to consider cyclic codes over $R$ of length $2n$ in this paper.

\vskip 3mm\par
   The present paper is organized as follows.
In Section 2, we provide an explicit representation for each cyclic code over $R$ of length $2n$ and give a formula to count the number of codewords in each code. As a corollary, we obtain a formula for the number of all such cyclic codes. In Section 3, we determine the dual code of each cyclic code over $R$ of length $2n$ and investigate self-duality for such codes.
Finally, we list all $293687$ cyclic codes and all $791$ self-dual cyclic codes of length $14$ over $\mathbb{F}_2[u]/\langle u^4\rangle$ in Section 4.

\section{Representation for all distinct cyclic codes over $R$ of length $2n$}
In this section, we investigate the structures of $R[x]/\langle x^{2n}-1\rangle$ first. Then
give an explicit presentation for all distinct cyclic codes over $R$ of length $2n$.

\par
   As $n$ is odd, there are pairwise coprime monic irreducible polynomials $f_1(x),\ldots,f_r(x)$ in $\mathbb{F}_{2^m}[x]$
such that $x^n-1=f_1(x)\ldots f_r(x)$. Then we have
\begin{center}
$x^{2n}-1=(x^n-1)^2=f_1(x)^2\ldots f_r(x)^2$.
\end{center}
Let $1\leq j\leq r$. We assume ${\rm deg}(f_j(x))=d_j$ and denote $F_j(x)=\frac{x^{2n}-1}{f_j(x)^2}$. Then ${\rm gcd}(F_j(x),f_j(x)^2)=1$
and there exist $a_j(x),b_j(x)\in\mathbb{F}_{2^m}[x]$ such that
\begin{equation}
a_j(x)F_j(x)+b_j(x)f_j(x)^2=1.
\end{equation}
From now on, we denote ${\cal A}=\mathbb{F}_{2^m}[x]/\langle x^{2n}-1\rangle$,
${\cal K}_j=\mathbb{F}_{2^m}[x]/\langle f_j(x)^2\rangle$ and set
\begin{equation}
\varepsilon_j(x)\equiv a_j(x)F_j(x)=1-b_j(x)f_j(x)^2 \ ({\rm mod} \ x^{2n}-1).
\end{equation}
From classical ring theory and coding theory we deduce the following lemma.

\vskip 3mm
\noindent
  {\bf Lemma 2.1} \textit{Using the notations above, we have the following}:

\par
  (i) \textit{$\varepsilon_1(x)+\ldots+\varepsilon_r(x)=1$, $\varepsilon_j(x)^2=\varepsilon_j(x)$
and $\varepsilon_j(x)\varepsilon_l(x)=0$  in the ring ${\cal A}$ for all $1\leq j\neq l\leq r$}.

\par
  (ii) \textit{${\cal A}={\cal A}_1\oplus\ldots \oplus{\cal A}_r$ where ${\cal A}_j={\cal A}\varepsilon_j(x)$ with
$\varepsilon_j(x)$ as its multiplicative identity and satisfies ${\cal A}_j{\cal A}_l=\{0\}$ for all $1\leq j\neq l\leq r$}.

\par
  (iii) \textit{For any integer $j$, $1\leq j\leq r$, define $\varphi_j: a(x)\mapsto \varepsilon_j(x)a(x)$ $(${\rm mod} $x^{2n}-1)$
for any $a(x)\in {\cal K}_j$. Then $\varphi_j$ is a ring isomorphism from ${\cal K}_j$ onto ${\cal A}_j$ and can be extended to a ring
isomorphism from ${\cal K}_j[u]/\langle u^k\rangle$ onto ${\cal A}_j[u]/\langle u^k\rangle$ in the natural way that:
for any $a_0(x),a_1(x),\ldots, a_{k-1}(x)\in \mathcal{K}_j$,}
$$\varphi_j: \sum_{l=0}^{k-1}a_l(x)u^l\mapsto \sum_{l=0}^{k-1}\varphi_j(a_l(x))u^l=\varepsilon_j(x)\left(\sum_{l=0}^{k-1}a_l(x)u^l\right) \
({\rm mod} \ x^{2n}-1).$$

\par
  (iv) \textit{For any integer $j$, $1\leq j\leq r$, ${\cal A}_j$ is a cyclic code over $\mathbb{F}_{2^m}$
of length $2n$ with parity check polynomial $f_j(x)^2$}.

\vskip 3mm
\par
   For any $\alpha(x)=\sum_{i=0}^{2n-1}\alpha_ix^i\in R[x]/\langle x^{2n}-1\rangle$
where $\alpha_i=\sum_{j=0}^{k-1}a_{i,j}u^j\in R$ with $a_{i,j}\in \mathbb{F}_{2^m}$ for all
$i=0,1,\ldots,2n-1$ and $j=0,1,\ldots,k-1$, we have
$$\alpha(x)=(1,x,\ldots,x^{2n-1})\left(\begin{array}{cccc}a_{0,0} & a_{0,1} & \ldots & a_{0,k-1}
\cr a_{1,0} & a_{1,1} & \ldots & a_{1,k-1} \cr \ldots & \ldots & \ldots & \ldots \cr a_{2n-1,0} & a_{2n-1,1} & \ldots & a_{2n-1,k-1}\end{array}\right)\left(\begin{array}{c}1\cr u\cr \ldots \cr u^{k-1}\end{array}\right).$$
Now, define $\Psi(\alpha(x))=a_0(x)+a_1(x)u+\ldots+a_{k-1}(x)u^{k-1}\in {\cal A}[u]/\langle u^k\rangle$
where $a_j(x)=a_{0,j}+a_{1,j}x+\ldots+a_{2n-1,j}x^{2n-1}\in {\cal A}$ for all $j=0,1,\ldots,k-1$. Then by a direct
calculation, one can easily verify the following lemma.

\vskip 3mm
\noindent
  {\bf Lemma 2.2} \textit{The map $\Psi$ defined above is a ring isomorphism from $R[x]/\langle x^{2n}-1\rangle$ onto
${\cal A}[u]/\langle u^k\rangle={\cal A}+u{\cal A}+\ldots+u^{k-1}{\cal A}$ $(u^k=0)$}.

\vskip 3mm
\noindent
  {\bf Remark} In the rest of this paper, we will identify $R[x]/\langle x^{2n}-1\rangle$ with ${\cal A}[u]/\langle u^k\rangle$
under the ring isomorphism $\Psi$.

\par
   Therefore, in order to determine cyclic codes over $R$ of length $2n$, i.e., ideals of
$R[x]/\langle x^{2n}-1\rangle$, it is sufficient to determine ideals of the ring ${\cal A}[u]/\langle u^k\rangle$.
Precisely, we have the following theorem.

\vskip 3mm
\noindent
  {\bf Theorem 2.3} \textit{Using the notations above, the following statements are equivalent}:

\par
  (i) \textit{${\cal C}$ is a cyclic code over $R$ of length $2n$}.

\par
  (ii) \textit{For any integer $j$, $1\leq j\leq r$, there is a unique ideal ${\cal C}_j$ of
the ring ${\cal A}_j[u]/\langle u^k\rangle$ such that ${\cal C}=\oplus_{j=1}^r{\cal C}_j$}.

\par
  (iii) \textit{For any integer $j$, $1\leq j\leq r$, there is a unique ideal $C_j$ of
the ring ${\cal K}_j[u]/\langle u^k\rangle$ such that ${\cal C}=\oplus_{j=1}^r\varepsilon_j(x)C_j$
$({\rm mod} \ x^{2n}-1)$}.

\vskip 3mm
\noindent
  \textit{Proof} (i)$\Leftrightarrow$(ii). By Lemma 2.1(ii) ${\cal A}$ is a direct product of its
ideals ${\cal A}_1,\ldots, {\cal A}_r$, which implies
${\cal A}[u]/\langle u^k\rangle=\oplus_{j=1}^r{\cal A}_j[u]/\langle u^k\rangle$. Hence every ideal
${\cal C}$ of ${\cal A}[u]/\langle u^k\rangle$ is uniquely decomposed as ${\cal C}=\oplus_{j=1}^r{\cal C}_j$,
where  ${\cal C}_j$ is an ideal of ${\cal A}_j[u]/\langle u^k\rangle$.

\par
   (ii)$\Leftrightarrow$(iii). Let $1\leq j\leq r$. By Lemma 2.1(iii), for any ideal ${\cal C}_j$ of ${\cal A}_j[u]/\langle u^k\rangle$ there is a unique ideal
$C_j$ of  ${\cal K}_j[u]/\langle u^k\rangle$ such that ${\cal C}_j=\varphi_j(C_j)=\varepsilon_j(x)C_j$.
\hfill $\Box$

\vskip 3mm
\noindent
  {\bf Remark} Using the notations of Theorem 2.3, ${\cal C}=\oplus_{j=1}^r\varepsilon_j(x)C_j$, where $C_j$
is an ideal of ${\cal K}_j[u]/\langle u^k\rangle$ for $j=1,\ldots,r$, is called a
\textit{canonical form decomposition} of the cyclic code ${\cal C}$ over $R$ of length $2n$.

\par
  By Theorem 2.3, in order to determine ideals of ${\cal A}[u]/\langle u^k\rangle$,
it is sufficient to determine ideals of ${\cal K}_j[u]/\langle u^k\rangle$ for all $j=1,\ldots,r$.
Let $1\leq j\leq r$ and $1\leq s\leq k$. In the reset of this paper, we adopt the following notations:

\vskip 2mm\par
  $\bullet$ ${\cal F}_j=\mathbb{F}_{2^m}[x]/\langle f_j(x)\rangle=\{\sum_{l=0}^{d_j-1}a_lx^l
\mid a_0,a_1,\ldots,a_{d_j-1}\in \mathbb{F}_{2^m}\}$.

\vskip 2mm\par
  $\bullet$ ${\cal F}_j[u]/\langle u^s\rangle=\{\sum_{h=0}^{s-1}b_h(x)u^h\mid b_0(x),b_1(x),\ldots,b_{s-1}(x)\in {\cal F}_j\}
={\cal F}_j+u{\cal F}_j+\ldots+u^{s-1}{\cal F}_j$ ($u^s=0$).

\vskip 2mm\par
 As $f_j(x)$ is a monic irreducible polynomial in $\mathbb{F}_{2^m}[x]$ of degree $d_j$, the following lemma follows from polynomial theory over finite fields and finite chain ring theory (cf. [7])
immediately. Here, we omit its proof.

\vskip 3mm\noindent
  {\bf Lemma 2.4} \textit{Using the notations above, for any $1\leq j\leq r$ and $1\leq s\leq k$ we have the following}:

\par
  (i) \textit{${\cal F}_j$ is an extension field of $\mathbb{F}_{2^m}$ with cardinality $2^{md_j}$}.

\par
  (ii) \textit{${\cal F}_j[u]/\langle u^s\rangle$
is a finite commutative chain ring with the unique maximal ideal $u({\cal F}_j[u]/\langle u^s\rangle)$,
the nilpotency index of $u$ is equal to $s$ and the residue field of ${\cal F}_j[u]/\langle u^s\rangle$ is $({\cal F}_j[u]/\langle u^s\rangle)/u({\cal F}_j[u]/\langle u^s\rangle)
\cong {\cal F}_j$}.

\par
  (iii) \textit{Every element $\alpha$ of ${\cal F}_j[u]/\langle u^s\rangle$
has a unique $u$-adic expansion}:
\begin{center}
$\alpha=t_0(x)+ut_1(x)+\ldots+u^{s-1}t_{s-1}(x)$, $t_0(x),t_1(x),\ldots,t_{s-1}(x)\in {\cal F}_j$.
\end{center}

\par
  (iv) \textit{The group of invertible elements of ${\cal F}_j[u]/\langle u^s\rangle$ is given by
\begin{center}
$({\cal F}_j[u]/\langle u^s\rangle)^{\times}=\{\sum_{h=0}^{s-1}b_h(x)u^h\mid b_0(x)\neq 0, b_0(x),b_1(x),\ldots,b_{s-1}(x)\in {\cal F}_j\}$
\end{center}
and $|({\cal F}_j[u]/\langle u^s\rangle)^{\times}|=(2^{md_j}-1)2^{(s-1)md_j}$. Then for any
nonzero element $\alpha$ of ${\cal F}_j[u]/\langle u^s\rangle$ there is a unique
integer $e$, $0\leq e\leq s-1$, such that $\alpha=u^e\omega$ for some $\omega\in ({\cal F}_j[u]/\langle u^s\rangle)^{\times}$}.

\par
  (v) \textit{Denote ${\cal T}_j=\{\sum_{l=0}^{d_j-1}a_lx^l
\mid a_0,a_1,\ldots,a_{d_j-1}\in \mathbb{F}_{2^m}\}\subset {\cal K}_j$. Then the subset ${\cal T}_j$ of
${\cal K}_j$ satisfies the following properties}:

\noindent
$\diamond$ \textit{For any $a(x)\in {\cal K}_j$, there is a unique $t(x)\in {\cal T}_j$ such that $a(x)\equiv t(x)$ $({\rm mod} \ f_j(x))$
as polynomials in $\mathbb{F}_{2^m}[x]$};

\noindent
 $\diamond$ \textit{$t_1(x)\not\equiv t_2(x)$ $({\rm mod} \ f_j(x))$, for any $t_1(x),t_2(x)\in {\cal T}_j$ satisfying $t_1(x)\neq t_2(x)$}.

 \textit{Furthermore, ${\cal K}_j$ is a finite chain ring
with the maximal ideal $f_j(x){\cal K}_j$ where $f_j(x)^2=0$, and every element of ${\cal K}_j$ has a unique $f_j(x)$-adic expansion}:
\begin{center}
$t_0(x)+f_j(x)t_1(x)$, $t_0(x),t_1(x)\in {\cal T}_j$.
\end{center}

\par
  (vi) \textit{Every element $\vartheta$ of ${\cal K}_j[u]/\langle u^s\rangle$ can be uniquely expressed as}
\begin{center}
$\vartheta=\alpha_0+f_j(x)\alpha_1$, $\alpha_0,\alpha_1\in {\cal T}_j[u]/\langle u^s\rangle$,
\end{center}
\textit{where ${\cal T}_j[u]/\langle u^s\rangle=\{\sum_{l=0}^{s-1}t_l(x)u^l\mid t_{l}(x)\in {\cal T}_j, \
0\leq l\leq s-1\}
\subset {\cal K}_j[u]/\langle u^s\rangle$}.

\vskip 3mm\noindent
  {\bf Remark} ${\cal F}_j=\{\sum_{l=0}^{d_j-1}a_lx^l
\mid a_0,a_1,\ldots,a_{d_j-1}\in \mathbb{F}_{2^m}\}$ which is a finite field with operations defined
by the usual polynomial operations modulo $f_j(x)$. But ${\cal T}_j$ is only a subset of $\mathcal{K}_j$ since the
operations on $\mathcal{K}_j$ are defined by the usual polynomial operations modulo $f_j(x)^2$.
In spite of this, we have ${\cal F}_j={\cal T}_j$ as a set. Therefore,
${\cal F}_j[u]/\langle u^s\rangle=\{\sum_{i=0}^{s-1}\xi_iu^i\mid \xi_0,\xi_1,\ldots,\xi_{s-1}\in {\cal F}_j\}={\cal T}_j[u]/\langle u^s\rangle$
as a set. In this sense, ${\cal F}_j[u]/\langle u^s\rangle$ is not a subring but a subset of ${\cal K}_j[u]/\langle u^s\rangle$.

\vskip 3mm\par
   Now, define a map $\tau: a(x)\mapsto a(x)$ (mod $f_j(x)$) for all $a(x)\in {\cal K}_j=\mathbb{F}_{2^m}[x]/\langle f_j(x)^2\rangle$.
It is clear that $\tau$ is a surjective ring homomorphism from ${\cal K}_j$ onto ${\cal F}_j=\mathbb{F}_{2^m}[x]/\langle f_j(x)\rangle$
with kernel ${\rm Ker}(\tau)=f_j(x){\cal K}_j$. Then we extend $\tau$ to ${\cal K}_j[u]/\langle u^s\rangle={\cal K}_j+u{\cal K}_j+\ldots+u^{s-1}{\cal K}_j$ ($u^s=0$) by
\begin{center}
$\tau:\sum_{l=0}^{s-1}a_l(x)u^l\mapsto \sum_{l=0}^{s-1}\tau(a_l(x))u^l$ $(\forall a_0(x), a_1(x),\ldots,a_{s-1}(x)\in {\cal K}_j)$.
\end{center}
Then $\tau$ is a surjective ring homomorphism from ${\cal K}_j[u]/\langle u^s\rangle$ onto ${\cal F}_j[u]/\langle u^s\rangle$ with kernel ${\rm Ker}(\tau)=f_j(x)({\cal K}_j[u]/\langle u^s\rangle)=\{f_j(x)\xi\mid \xi\in {\cal T}_j[u]/\langle u^s\rangle\}$.

\par
    Let $\alpha\in {\cal F}_j[u]/\langle u^s\rangle$ and $f_j(x)\xi\in f_j(x)({\cal K}_j[u]/\langle u^s\rangle)$
where $\xi\in {\cal T}_j[u]/\langle u^s\rangle$. As ${\cal T}_j[u]/\langle u^s\rangle={\cal F}_j[u]/\langle u^s\rangle$ as a set,
we regard $\xi$ as an element of ${\cal F}_j[u]/\langle u^s\rangle$ and calculate the product $\xi\alpha$ of
$\xi$ and $\alpha$ in the ring ${\cal F}_j[u]/\langle u^s\rangle$. Also, we can regard $\xi\alpha$ as an
element of ${\cal T}_j[u]/\langle u^s\rangle$. Then by Lemma 2.4(vi), we obtain a unique element $f_j(x)(\xi\alpha)$ in $f_j(x)({\cal K}_j[u]/\langle u^s\rangle)$. Now, we define
\begin{center}
$\alpha\cdot f_j(x)\xi=f_j(x)(\xi\alpha)\in f_j(x)({\cal K}_j[u]/\langle u^s\rangle)$.
\end{center}
Especially, we have $\alpha\cdot f_j(x)=f_j(x)\alpha\in f_j(x)({\cal K}_j[u]/\langle u^s\rangle)$ if $\xi=1$.
Then $f_j(x)({\cal K}_j[u]/\langle u^s\rangle)$ is an ${\cal F}_j[u]/\langle u^s\rangle$-module. Moreover, we have

\vskip 3mm
\noindent
   {\bf Lemma 2.5} \textit{With the scalar multiplication defined above, the map
$\psi_j$ defined by $\psi_j(\alpha)=f_j(x)\alpha\in {\cal K}_j[u]/\langle u^s\rangle$ $(\forall \alpha \in{\cal F}_j[u]/\langle u^s\rangle)$ is an
${\cal F}_j[u]/\langle u^s\rangle$-module isomorphism from ${\cal F}_j[u]/\langle u^s\rangle$ onto
$f_j(x)({\cal K}_j[u]/\langle u^s\rangle)$}.

\vskip 3mm
\noindent
   \textit{Proof} By Lemma 2.4(vi), every element $\vartheta$ of ${\cal K}_j[u]/\langle u^s\rangle$ can be uniquely
expressed as $\vartheta=\alpha_0+f_j(x)\alpha_1$ where $\alpha_0,\alpha_1\in {\cal T}_j[u]/\langle u^s\rangle={\cal F}_j[u]/\langle u^s\rangle$.
From this and by $f_j(x)^2=0$ in ${\cal K}_j[u]/\langle u^s\rangle$, we deduce that
$f_j(x)\vartheta=f_j(x)\alpha_0$. Hence $\psi_j$ is a bijection. Moreover, for any $\alpha_1,\alpha_2\in {\cal F}_j[u]/\langle u^s\rangle$
we assume that $\alpha_1+\alpha_2=\beta$ and $\alpha_1\alpha_2=\gamma$ in ${\cal F}_j[u]/\langle u^s\rangle$. Now, we regard
$\alpha_1$ and $\alpha_2$ as elements of ${\cal K}_j[u]/\langle u^s\rangle$ and assume
$\alpha_1+\alpha_2=\zeta$ and $\alpha_1\alpha_2=\eta$ in ${\cal K}_j[u]/\langle u^s\rangle$.
Since $\tau$ is a surjective ring homomorphism from ${\cal K}_j[u]/\langle u^s\rangle$ onto ${\cal F}_j[u]/\langle u^s\rangle$,
by the definition of $\tau$ we have $\tau(\zeta)=\alpha_1+\alpha_2=\beta$ and $\tau(\eta)=\alpha_1\alpha_2=\gamma$, which
imply $\zeta=\beta+f_j(x)\beta_1$ and $\eta=\gamma+f_j(x)\gamma_1$ for some $\beta_1,\gamma_1\in {\cal T}_j[u]/\langle u^s\rangle$. By $f_j(x)^2=0$
in ${\cal K}_j[u]/\langle u^s\rangle$,
we have
\begin{eqnarray*}
\psi_j(\alpha_1+\alpha_2)&=&f_j(x)\beta=f_j(x)(\beta+f_j(x)\beta_1)=f_j(x)\zeta=f_j(x)(\alpha_1+\alpha_2)\\
   &=&f_j(x)\alpha_1+f_j(x)\alpha_2=\psi_j(\alpha_1)+\psi_j(\alpha_2)
\end{eqnarray*}
 and $\psi_j(\alpha_1\alpha_2)=f_j(x)\gamma=f_j(x)(\gamma+f_j(x)\gamma_1)=f_j(x)\eta=f_j(x)(\alpha_1\alpha_2)
=\alpha_1(f_j(x)\alpha_2)=\alpha_1\psi_j(\alpha_2)$. Hence $\psi_j$ is an ${\cal F}_j[u]/\langle u^s\rangle$-module isomorphism from ${\cal F}_j[u]/\langle u^s\rangle$ onto
$f_j(x)({\cal K}_j[u]/\langle u^s\rangle)$.
\hfill $\Box$

\vskip 3mm
\noindent
   {\bf Remark} By Lemma 2.5, every element of $f_j(x)({\cal K}_j[u]/\langle u^s\rangle)$ can be
uniquely expressed as: $f_j(x)\alpha=\psi_j(\alpha)$, $\alpha\in {\cal F}_j[u]/\langle u^s\rangle$, where
we regard $\alpha$ in the expression $f_j(x)\alpha$ as an element of ${\cal T}_j[u]/\langle u^s\rangle$
in the rest of this paper.

\vskip 3mm\par
  Then we present all distinct ideals
of the ${\cal K}_j[u]/\langle u^k\rangle$.

\vskip 3mm
\noindent
  {\bf Theorem 2.6} \textit{Using the notations above, all distinct ideals of ${\cal K}_j[u]/\langle u^k\rangle$ are
given by the following table}:
{\small\begin{center}
\begin{tabular}{llll}\hline
case &  number of ideals  &  $C_j$ (ideal of ${\cal K}_j[u]/\langle u^k\rangle$)    &   $|C_j|$ \\ \hline
I.   & $k+1$  & $\bullet$ $\langle u^i\rangle$ \ $(0\leq i\leq k)$ & $2^{2md_j(k-i)}$ \\
II.   & $k$     & $\bullet$  $\langle u^s f_j(x)\rangle$ \ $(0\leq s\leq k-1)$ &  $2^{md_j(k-s)}$  \\
III.   & $\Omega_1(2^{md_j},k)$ & $\bullet$   $\langle u^i+u^tf_j(x)\omega\rangle$ &  $2^{2md_j(k-i)}$ \\
     &                            & \ \ ($\omega\in ({\cal F}_j[u]/\langle u^{i-t}\rangle)^{\times}$, $t\geq 2i-k$, & \\
     &                            & \ \  $ 0\leq t<i\leq k-1)$                                 &              \\
IV.  & $\Omega_2(2^{md_j},k)$     & $\bullet$   $\langle u^i+u^tf_j(x)\omega \rangle$ &  $2^{md_j(k-t)}$ \\
     &                            & \ \  ($\omega\in ({\cal F}_j[u]/\langle u^{k-i}\rangle)^{\times}$, $t< 2i-k$, & \\
     &                            & \ \  $ 0\leq t<i\leq k-1)$        \\
V.   &  $\frac{1}{2}k(k-1)$    & $\bullet$    $\langle u^i,u^sf_j(x)\rangle$ &  $2^{md_j(2k-(i+s))}$ \\
     &                         &  $ \ \ (0\leq s<i\leq k-1)$ & \\
VI.   &  $(2^{md_j}-1)$    &  $\bullet$   $\langle u^i+u^tf_j(x)\omega, u^sf_j(x)\rangle$ &  $2^{md_j(2k-(i+s))}$ \\
     & $\cdot \Gamma(2^{md_j},k)$ & \ \  $(\omega\in ({\cal F}_j[u]/\langle u^{s-t}\rangle)^{\times}$,  $i+s\leq k+t-1$, & \\
     &                         & \ \ $0\leq t<s<i\leq k-1)$ & \\ \hline
\end{tabular}
\end{center}}

\noindent
\textit{where}

\vskip 2mm\noindent
  $\diamond$ $\Omega_1(2^{md_j},k)=\left\{\begin{array}{ll}\frac{2^{md_j(\frac{k}{2}+1)}+
2^{md_j\cdot\frac{k}{2}}-2}{2^{md_j}-1}-(k+1), & {\rm if} \ k \ {\rm is} \ {\rm even};\cr
\frac{2(2^{md_j\cdot\frac{k+1}{2}}-1)}{2^{md_j}-1}-(k+1), & {\rm if} \ k \ {\rm is} \ {\rm odd}.\end{array}\right.$

\vskip 2mm\noindent
  $\diamond$ $\Omega_2(2^{md_j},k)=\left\{\begin{array}{ll}(2^{md_j}-1)\sum_{i=\frac{k}{2}+1}^{k-1}(2i-k)2^{md_j(k-i-1)}, & {\rm if} \ k \ {\rm is} \ {\rm even};\cr
(2^{md_j}-1)\sum_{i=\frac{k+1}{2}}^{k-1}(2i-k)2^{md_j(k-i-1)}, & {\rm if} \ k \ {\rm is} \ {\rm odd}.\end{array}\right.$

\vskip 2mm\noindent
  $\diamond$ \textit{$\Gamma(2^{md_j},k)$ can be calculated by the following recurrence formula}:

\vskip 2mm\par
   \textit{$\Gamma(2^{md_j},\rho)=0$ for $\rho=1,2,3$, $\Gamma(2^{md_j},\rho)=1$ for $\rho=4$};

\vskip 2mm\par
  \textit{$\Gamma(2^{md_j},\rho)=\Gamma(2^{md_j},\rho-1)+\sum_{s=1}^{\lfloor\frac{\rho}{2}\rfloor-1}(\rho-2s-1)2^{md_j(s-1)}$ for $\rho\geq 5$}.

\vskip 2mm \noindent
  \textit{Therefore, the number of all
distinct ideals of the ring  ${\cal K}_j[u]/\langle u^k\rangle$ is equal to}
\begin{center}
$N_{(2^m,d_j,k)}=1+\frac{k(k+3)}{2}+\Omega_1(2^{md_j},k)+\Omega_2(2^{md_j},k)+(2^{md_j}-1)\Gamma(2^{md_j},k)$.
\end{center}

\vskip 3mm
\noindent
   \textit{Proof} In order to simplify the notations, we denote $\Upsilon_j={\cal K}_j[u]/\langle u^k\rangle$
and ${\cal R}_j={\cal F}_j[u]/\langle u^k\rangle$. By Lemma 2.4(ii), ${\cal R}_j$ is a finite chain ring with maximal
ideal $u{\cal R}_j$, the nilpotency index of $u$ is equal to $k$ and the residue field of ${\cal R}_j$ is
${\cal R}_j/(u{\cal R}_j)\cong{\cal F}_j$. Then from finite chain ring theory (cf. [7]), we deduce that all ideals of ${\cal R}_j$ is given by $u^l{\cal R}_j$, where $0\leq l\leq k$,
and $|u^l{\cal R}_j|=|{\cal F}_j|^{k-l}=2^{md_j(k-l)}$. Moreover, we have $f_j(x)\Upsilon_j=f_j(x){\cal R}_j$ by Lemma 2.5.

\par
   Let $C$ be any ideal of $\Upsilon_j$. We define
\begin{center}
${\rm Tor}_0(C)=\tau(C)$, ${\rm Tor}_1(C)=\tau(\{\alpha\in \Upsilon_j\mid f_j(x)\alpha\in C\})$.
\end{center}
Then it can be easily verify that both ${\rm Tor}_0(C)$ and ${\rm Tor}_1(C)$ are ideals of ${\cal R}_j$
satisfying ${\rm Tor}_0(C)\subseteq {\rm Tor}_1(C)$. Hence there is a unique pair $(i,s)$ of integers, $0\leq s\leq i\leq k$,
such that ${\rm Tor}_0(C)=u^i{\cal R}_j$ and ${\rm Tor}_1(C)=u^s{\cal R}_j$.

\par
  Let $\tau|_C$ be the restriction of $\tau$ to $C$. Then $\tau|_C$
is a surjective ring homomorphism from $C$ onto ${\rm Tor}_0(C)$ with kernel ${\rm Ker}(\tau|_C)
=f_j(x){\rm Tor}_1(C)$. As $|{\rm Ker}(\tau|_C)|=|{\rm Tor}_1(C)|$, by the ring isomorphism theorems it follows
that
\begin{equation}
|C|=|{\rm Tor}_0(C)||{\rm Tor}_1(C)|=|u^i{\cal R}_j||u^s{\cal R}_j|=2^{md_j(2k-(i+s))}.
\end{equation}
Then we have the following cases.

\vskip 2mm\par
  Case (i) $s=i$. In this case, we have $|C|=2^{2md_j(k-i)}$ by Equation (3).

\par
  When $s=k$, then $i=k$ and $C=\{0\}=u^k\Upsilon_j=\langle u^k\rangle$.

\par
  Let $0\leq i\leq k-1$.
  By $u^i\in u^i{\cal R}_j={\rm Tor}_0(C)$, there exists $\alpha\in \Upsilon_j$ such that
$u^i+f_j(x)\alpha\in C$. Then by Lemma 2.5 we can assume that $\alpha\in {\cal R}_j$.

\par
  It is obvious that $\langle u^i+f_j(x)\alpha\rangle=\Upsilon_j(u^i+f_j(x)\alpha)\subseteq C$.
Conversely, let $\xi\in C$. By ${\rm Tor}_0(C)=u^i{\cal R}_j$ and Lemma 2.5 we have that
$\xi=u^i\beta+f_j(x)\gamma$ for some $\beta,\gamma \in {\cal R}_j$, which implies
$f_j(x)(\gamma-\alpha\beta)=\xi-(u^i+f_j(x)\alpha)\beta\in C$. By Lemma 2.5 we may regard $\gamma-\alpha\beta$
as an element of ${\cal R}_j$. Hence $\gamma-\alpha\beta=\tau(\gamma-\alpha\beta)\in {\rm Tor}_1(C)=u^i{\cal R}_j$,
which implies that $\gamma-\alpha\beta=u^i\delta$ for some $\delta\in {\cal R}_j$. Therefore, we
have $\xi=(u^i+f_j(x)\alpha)\beta+(u^i+f_j(x)\alpha)f_j(x)\delta=(u^i+f_j(x)\alpha)(\beta+f_j(x)\delta)\in
\langle u^i+f_j(x)\alpha\rangle$. Hence $C=\langle u^i+f_j(x)\alpha\rangle$.

\par
  When $i=0$, $1+f_j(x)\alpha\in C$. By $(1+f_j(x)\alpha)(1-f_j(x)\alpha)=1-f_j(x)^2\alpha^2=1$ in
$\Upsilon_j$, we see that $1+f_j(x)\alpha$ is an invertible element of $\Upsilon_j$, which implies
$C=\Upsilon_j=\langle u^0\rangle$. In the following, we assume $1\leq i\leq k-1$.

\par
  Obviously, we have $f_j(x)u^i=f_j(x)(u^i+f_j(x)\alpha)\in C$. By Lemma 2.4(iv), we can write $\alpha$ as $\alpha=u^i\beta$ or $\alpha=u^t\omega+u^i\beta$ for some $\beta\in {\cal R}_j$, $0\leq t\leq i-1$
and $\omega\in {\cal R}_j^{\times}$.

\par
   When $\alpha=u^i\beta$, we have
$u^i=u^i+f_j(x)\alpha-(f_j(x)u^i)\beta \in C$, and so $C=\langle u^i\rangle$.

\par
  Let $\alpha=u^t\omega+u^i\beta$. Then we have $u^i+u^tf_j(x)\omega=u^i+f_j(x)\alpha-(f_j(x)u^i)\beta \in C$,
and hence $C=\langle u^i+u^tf_j(x)\omega\rangle$. In this case, we have $u^{k-i+t}f_j(x)=
u^{k-i}\omega^{-1}(u^i+u^tf_j(x)\omega)\in C$ where $\omega^{-1}$ is the inverse of $\omega$
in ${\cal R}_j$. Therefore, $u^{k-i+t}\in {\rm Tor}_1(C)=u^i{\cal R}_j$, which implies
$k-i+t\geq i$, and hence $t\geq 2i-k$.

\par
   Now, let $\omega_1,\omega_2\in {\cal R}_j^{\times}$ and $0\leq t_1,t_2\leq k-1$ satisfying
$C=\langle u^i+u^{t_1}f_j(x)\omega_1\rangle=\langle u^i+u^{t_2}f_j(x)\omega_2\rangle$.
Then $f_j(x)(u^{t_1}\omega_1-u^{t_2}\omega_2)=(u^i+u^{t_1}f_j(x)\omega_1)$ $-(u^i+u^{t_2}f_j(x)\omega_2)\in C$,
which implies $u^{t_1}\omega_1-u^{t_2}\omega_2\in {\rm Tor}_1(C)=u^i{\cal R}_j$. From this we deduce
that $t_1=t_2=t$ and $u^t(\omega_1-\omega_2)\in u^i{\cal R}_j$. Then by finite chain ring theory (cf. [7])
and $u^t(\omega_1-\omega_2)\in u^i{\cal R}_j$, it follows that $\omega_1\equiv\omega_2$ (mod $u^{i-t}$), i.e.,
$\omega_1=\omega_2$ as elements of $({\cal R}_j/(u^{i-t}{\cal R}_j)^{\times}$. Finally, by ${\cal R}_j={\cal F}_j[u]/\langle u^k\rangle$
we have $({\cal R}_j/(u^{i-t}{\cal R}_j)^{\times}={\cal F}_j[u]/\langle u^{i-t}\rangle$ up to a natural ring isomorphism.

\par
  As stated above, we conclude that all distinct ideals $C$ of $\Upsilon_j$ is given by (I) and (III) in the table.

\vskip 2mm\par
  Case (ii) $i=k$ and $0\leq s\leq k-1$.

\par
  In this case, we have ${\rm Tor}_0(C)=\{0\}$
and $|C|=2^{md_j(2k-(k+s))}=2^{md_j(k-s)}$ by Equation (3). Moreover, by ${\rm Tor}_1(C)=u^s {\cal R}_j$ it can be easily verified that all distinct ideals of $\Upsilon_j$ are given by: (II) $C=\langle u^sf_j(x)\rangle$ where $0\leq s\leq k-1$.

\vskip 2mm\par
  Case (iii) $s=0$ and $1\leq i\leq k-1$.

\par
  In this case, we have $|C|=2^{md_j(2k-i)}$ by Equation (3).
Moreover, by $1\in {\rm Tor}_1(C)=u^0 {\cal R}_j$ we conclude that $f_j(x)=u^0f_j(x)\in C$. Then by ${\rm Tor}_0(C)=u^i {\cal R}_j$ it follows that
$C=\langle u^i,f_j(x)\rangle$ immediately.

\vskip 2mm\par
   Case (iv) $1\leq s<i\leq k-1$. In this case, $|C|=2^{md_j(2k-(i+s))}$ by (3).

\par
   By ${\rm Tor}_0(C)=u^i {\cal R}_j$ and
${\rm Tor}_1(C)=u^s {\cal R}_j$ we have $u^sf_j(x)\in C$ and there exists $\alpha\in {\cal R}_j$ such
that $u^i+f_j(x)\alpha \in C$, which implies $\langle u^i+f_j(x)\alpha, u^s f_j(x)\rangle\subseteq C$.
Conversely, let $\xi\in C$. By ${\rm Tor}_0(C)=u^i {\cal R}_j$ there exist $\beta,\gamma\in {\cal R}_j$ such that $\xi=u^i\beta+f_j(x)\gamma$.
Then from $f_j(x)(\gamma-\alpha\beta)=\xi-(u^i+f_j(x)\alpha)\beta\in C$ where $\gamma-\alpha\beta\in {\cal R}_j$, we deduce
$\gamma-\alpha\beta\in {\rm Tor}_1(C)=u^s {\cal R}_j$, which implies $\gamma-\alpha\beta=u^s \delta$ for some $\delta\in {\cal R}_j$,
and so $\xi=(u^i+f_j(x)\alpha)\beta+(f_j(x)u^s)\delta\in \langle u^i+f_j(x)\alpha, f_j(x)u^s\rangle$.
Hence $C=\langle u^i+f_j(x)\alpha, u^sf_j(x)\rangle$. Furthermore,  by $u^sf_j(x)\in C$ and an argument similar to the proof of Case (i)
we can assume that $\alpha=0$ or $\alpha=u^t\omega$, where $\omega\in {\cal R}_j^{\times}$ and $0\leq t\leq s-1$.

\par
  (iv-1) When $\alpha=0$, we have $C=\langle u^i, u^sf_j(x)\rangle$ given by (V) in the table.

\par
  (iv-2) Let $C=\langle u^i+u^tf_j(x)\omega, u^sf_j(x)\rangle$ where $\omega\in {\cal R}_j^{\times}$ and $0\leq t\leq s-1$.

\par
  Assume $C=\langle u^i+u^{t_1}f_j(x)\omega_1, u^sf_j(x)\rangle=\langle u^i+u^{t_2}f_j(x)\omega_2, u^sf_j(x)\rangle$
where $0\leq t_1,t_2\leq s-1$ and $\omega_1,\omega_2\in {\cal R}^{\times}$. Then $f_j(x)(u^{t_1}\omega_1-u^{t_2}\omega_2)=(u^i+u^{t_1}f_j(x)\omega_1)-(u^i+u^{t_2}f_j(x)\omega_2)\in C$, which
implies $u^{t_1}\omega_1-u^{t_2}\omega_2\in {\rm Tor}_1(C)=u^s {\cal R}_j$. From this and by an argument similar to the proof
of Case (i), we deduce that $t_1=t_2=t$ and $\omega_1\equiv \omega_2$ (mod $u^{s-t} {\cal R}_j$).
By the latter condition, we have $\omega=\omega_1=\omega_2\in ({\cal R}_j/(u^{s-t}{\cal R}_j))^{\times}$.
From this and by ${\cal R}_j/(u^{s-t}{\cal R}_j)={\cal F}_j[u]/\langle u^{s-t}\rangle$, we deduce
that all distinct ideals of $\Upsilon_j$ are given by: $C=\langle u^i+u^tf_j(x)\omega, u^sf_j(x)\rangle$ where $\omega\in ({\cal F}_j[u]/\langle u^{s-t}\rangle)^{\times}$
in this case.

\par
  As $f_j(x)u^{k-i+t}=u^{k-i}\omega^{-1}(u^i+u^tf_j(x)\omega)\in C$, we have
$u^{k-i+t}\in {\rm Tor}_1(C)=u^s {\cal R}_j$, which implies  $k-i+t\geq s$. So we have one of the following
two cases:

\par
  ($\diamondsuit$-1) $k-i+t=s$, i.e., $s-t=k-i$.

\par
  In this case, by $f_j(x)u^s=u^{k-i}\omega^{-1}(u^i+u^tf_j(x)\omega)$ we have
$C=\langle u^i+u^tf_j(x)\omega\rangle$ and $2i>i+s=k+t$, i.e., $t<2i-k$. Furthermore, we have
$\omega\in ({\cal F}_j[u]/\langle u^{k-i}\rangle)^{\times}$ and $|C|=2^{md_j(2k-(i+s))}
=2^{md_j(2k-(k+t))}=2^{md_j(k-t)}$. Hence $C$ is given by (IV) in the table.

\par
  ($\diamondsuit$-2) $k-i+t>s$, i.e., $i+s\leq k+t-1$. In this case, $C$ is given by (VI) in the table.

\par
  As stated above, we conclude that all distinct ideals and the number of elements in each ideal of $\Upsilon_j$ are given by
(I)--(VI) of the table.

\par
   It is obvious that the number of ideals in (I), (II) and (V) is equal to $k+1$, $k$ and $\frac{1}{2}k(k-1)$ respectively. Then we count the number of ideals in (III), (IV) and (V) respectively.
In order to simplify the notation, we denote $q_j=2^{md_j}$ in the following. Then by Lemma 2.4(iv) we know that
$({\cal F}_j[u]/\langle u^{l}\rangle)^{\times}=(q_j-1)q_j^{l-1}$ for all $l=1,\ldots,k$.

\par
   First, we count the number of ideals in (III).
Let $0\leq t<i\leq k-1$. Then we have one of the following two cases:

\par
  ($\diamondsuit$-1) $k$ is even. In this case, $t\geq 2i-k$ if and only if $i$ and $t$ satisfy one of the following two conditions:

\par
  ($\diamondsuit$-1-1) $i\leq \frac{k}{2}$, i.e., $2i\leq k$, and $0\leq t\leq i-1$. In this case,
the number of ideals is equal to
$N_1=\sum_{i=1}^{\frac{k}{2}}\sum_{t=0}^{i-1}(q_j-1)q_j^{i-t-1}=\frac{q_j^{\frac{k}{2}+1}-1}{q_j-1}-(\frac{k}{2}+1)$.

\par
  ($\diamondsuit$-1-2) $i\geq \frac{k}{2}+1$, i.e., $2i> k$, and $2i-k\leq t\leq i-1$.
In this case,
the number of ideals is equal to
$N_2=\sum_{i=\frac{k}{2}+1}^{k-1}\sum_{t=2i-k}^{i-1}(q_j-1)q_j^{i-t-1}
=\frac{q_j^{\frac{k}{2}}-1}{q_j-1}-\frac{k}{2}$.

\par
  Therefore, the number of ideals in (III) is equal to
  $\Omega_1(q_j,k)=N_1+N_2=\frac{q_j^{\frac{k}{2}+1}+
q_j^{\frac{k}{2}}-2}{q_j-1}-(k+1)$, where $k$ is even.

 ($\diamondsuit$-2) $k$ is odd. In this case, $t\geq 2i-k$ if and only if $i$ and $t$ satisfy one of the following two conditions:

\par
  ($\diamondsuit$-2-1) $i\leq \frac{k-1}{2}$, i.e., $2i\leq k$, and $0\leq t\leq i-1$. Then
the number of ideals is equal to
$N_1=\sum_{i=1}^{\frac{k-1}{2}}\sum_{t=0}^{i-1}(q_j-1)q_j^{i-t-1}=\frac{q_j^{\frac{k+1}{2}}-1}{q_j-1}-\frac{k+1}{2}$.

\par
  ($\diamondsuit$-2-2) $i\geq \frac{k-1}{2}+1$, i.e., $2i> k$, and $2i-k\leq t\leq i-1$.
Then
the number of ideals is equal to
$N_2=\sum_{i=\frac{k-1}{2}+1}^{k-1}\sum_{t=2i-k}^{i-1}(q_j-1)q_j^{i-t-1}
=\frac{q_j^{\frac{k+1}{2}}-1}{q_j-1}-\frac{k+1}{2}$.

\par
  Therefore, the number of ideals in (III) is equal to
  $\Omega_1(q_j,k)=N_1+N_2=\frac{2q_j^{\frac{k+1}{2}}-2}{q_j-1}-(k+1)$, where $k$ is odd.

\par
   Next we count the number of ideals in (IV).
Let $0\leq t<i\leq k-1$. Then we have one of the following two cases:

\par
  ($\diamondsuit$-1) $k$ is even. Then $t<2i-k$ if and only if $2i>k$, i.e., $i\geq \frac{k}{2}+1$,
and $0\leq t\leq 2i-k-1$. Hence the number of ideals in (IV) is equal to
  $\Omega_2(q_j,k)=\sum_{i=\frac{k}{2}+1}^{k-1}\sum_{t=0}^{2i-k-1}
(q_j-1)q_j^{k-i-1}
=(q_j-1)\sum_{i=\frac{k}{2}+1}^{k-1}(2i-k)q_j^{k-i-1}$.

\par
  ($\diamondsuit$-2) $k$ is odd. Then $t<2i-k$ if and only if $2i>k$, i.e., $i\geq \frac{k+1}{2}$,
and $0\leq t\leq 2i-k-1$. Hence the number of ideals in (IV) is equal to
  $\Omega_2(q_j,k)=\sum_{i=\frac{k+1}{2}}^{k-1}\sum_{t=0}^{2i-k-1}
(q_j-1)q_j^{k-i-1}
=(q_j-1)\sum_{i=\frac{k+1}{2}}^{k-1}(2i-k)q_j^{k-i-1}$.

\par
   Finally, denote $\Gamma(q_j,k)=\sum_{t=0}^{k-4}\sum_{t+1\leq s\leq \frac{k+t}{2}-1}\sum_{s+1\leq i\leq k+t-1-s}q_j^{s-t-1}$. Then
the number of ideals in (VI) is equal to $(q_j-1)\Gamma(q_j,k)$, where
$$
\Gamma(q_j,k)=\sum_{t=1}^{k-4}\sum_{t+1\leq s\leq \frac{k+t}{2}-1}\sum_{s+1\leq i\leq k+t-1-s}q_j^{s-t-1}
   +\sum_{1\leq s\leq \frac{k}{2}-1}\sum_{s+1\leq i\leq k-1-s}q_j^{s-1}$$
\begin{eqnarray*}
   &=&\sum_{t^{\prime}=0}^{(k-1)-4}\sum_{t^{\prime}+1\leq s^{\prime}\leq \frac{(k-1)+t^{\prime}}{2}-1}\sum_{s^{\prime}+1\leq i^{\prime}\leq (k-1)+t^{\prime}-1-s^{\prime}}q_j^{s^{\prime}-t^{\prime}-1} \ \ \ \ \ \ \  \\
   &&+\sum_{s=1}^{\lfloor \frac{k}{2}\rfloor-1}(k-2s-1)q_j^{s-1}\\
   &=&\Gamma(q_j,k-1)+\sum_{s=1}^{\lfloor \frac{k}{2}\rfloor-1}(k-2s-1)q_j^{s-1}
\end{eqnarray*}
when $k\geq 5$.

\par
   If $1\leq k\leq 3$, there is no triple $(t,s,i)$ of integers satisfying $0\leq t<s<i\leq k-1$ and $i+s\leq k+t-1$.
In this case, the number of ideals in (VI) is equal to $0$. Then we set $\Gamma(q_j,k)=0$ for $k=1,2,3$.

\par
   If $k=4$, there is a
unique triple $(t,s,i)=(0,1,2)$ of integers satisfying $0\leq t<s<i\leq k-1$ and $i+s\leq k+t-1$. In this case, all distinct ideals in (VI) are
given by $\langle u^2+f_j(x)\omega,uf_j(x)\rangle$, where $\omega\in ({\cal F}_j[u]/\langle u\rangle)^{\times}={\cal F}_j^{\times}$ and
$|{\cal F}_j^{\times}|=q_j-1$. Then
we set $\Gamma(q_j,4)=1$.

\par
  Therefore, the number of ideals of $\Upsilon_j$ is equal to $k+1+k+\Omega_1(q_j,k)+\Omega_2(q_j,k)
+\frac{1}{2}k(k-1)+(q_j-1)\Gamma(q_j,k)
=1+\frac{1}{2}k(k+3)+\Omega_1(q_j,k)+\Omega_2(q_j,k)+(q_j-1)\Gamma(q_j,k)$.
\hfill $\Box$

\vskip 3mm \noindent
   {\bf Corollary 2.7} \textit{Using the notations in Theorems 2.3 and 2.6, the number of cyclic codes over $\mathbb{F}_{2^m}[u]/\langle u^k\rangle$
of length $2n$ is equal to $\prod_{j=1}^rN_{(2^m,d_j,k)}$}.

\vskip 3mm \noindent
   \textit{Proof} It follows from Theorems 2.3 and 2.6 immediately.
\hfill $\Box$

\section{Dual codes and self-duality of cyclic codes over $R$ of length $2n$}
\noindent
   In this section, we give the dual code of each cyclic code over the ring $R=\mathbb{F}_{2^m}[u]/\langle u^k\rangle$
of length $2n$ where $n$ is odd, and then investigate the self-duality of these codes.

\par
   For any $\alpha=(\alpha_0,\alpha_1,\ldots,\alpha_{2n-1}), \beta=(\beta_0,\beta_1,\ldots,\beta_{2n-1})\in R^{2n}$,
where $\alpha_j,\beta_j\in R$ and $0\leq j\leq 2n-1$, recall that
the usual \textit{Euclidian inner product} of $\alpha$ and $\beta$ are defined by
$[\alpha,\beta]_E=\sum_{j=0}^{2n-1}\alpha_j\beta_j\in R$.
It is known that $[-,-]_E$ is a symmetric and non-degenerate bilinear form on the $R$-module
$R^{2n}$. Let $C$ be a linear code over $R$ of length ${2n}$, i.e.,
an $R$-submodule of $R^{2n}$. The \textit{Euclidian dual code}
of $C$ is defined by $C^{\bot_E}=\{\alpha\in R^{2n}\mid [\alpha,\beta]_E=0, \ \forall
\beta\in C\}$, and $C$ is said to be \textit{self-dual} if $C=C^{\bot_E}$. As usual, we will identify $\alpha\in R^{2n}$
with $\alpha(x)=\sum_{j=0}^{2n-1}\alpha_jx^j\in R[x]/\langle x^{2n}-1\rangle$ in this paper.
Then we define
$$\varrho(\alpha(x))=\alpha(x^{-1})=\alpha_0+\sum_{j=1}^{2n-1}\alpha_jx^{2n-j}, \ \forall \alpha(x)\in R[x]/\langle x^{2n}-1\rangle.$$
It is clear that $\varrho$ is a ring automorphism of $R[x]/\langle x^{2n}-1\rangle$ satisfying
$\varrho^{-1}=\varrho$. Now, by a direct calculation we get the following lemma.

\vskip 3mm \noindent
  {\bf Lemma 3.1} \textit{Let $\alpha,\beta\in R^{2n}$.
Then $[\alpha,\beta]_E=0$  if $\alpha(x)\varrho(\beta(x))=0$
in the ring $R[x]/\langle x^{2n}$ $-1\rangle$}.

\vskip 3mm \par
   By Lemma 2.2 and the remark followed this lemma, we have $R[x]/\langle x^{2n}-1\rangle=
{\cal A}[u]/\langle u^k\rangle$ where ${\cal A}=\mathbb{F}_{2^m}[x]/\langle x^{2n}-1\rangle$. It is obvious that the restriction of $\varrho$
to ${\cal A}$ is a ring automorphism of ${\cal A}$. We still denote this automorphism by $\varrho$, i.e.,
$\varrho(a(x))=a(x^{-1})$ for any $a(x)\in {\cal A}$.

\par
   Let $1\leq j\leq r$. By Equations (1) and (2) in Section 2, we have
\begin{equation}
\varrho(\varepsilon_j(x))=a_j(x^{-1})F_j(x^{-1})=1-b_j(x^{-1})f_j(x^{-1})^2 \ {\rm in} \ {\cal A}.
\end{equation}
For any polynomial $f(x)=\sum_{l=0}^da_lx^l\in \mathbb{F}_{2^m}[x]$ of degree $d\geq 1$, recall that
the \textit{reciprocal polynomial} of $f(x)$ is defined as $\widetilde{f}(x)=x^df(\frac{1}{x})=\sum_{l=0}^da_lx^{d-l}$, and
 $f(x)$ is said to be \textit{self-reciprocal} if $\widetilde{f}(x)=ef(x)$ for some $e\in \mathbb{F}_{2^m}^{\times}$. Then by Equation (1) in Section 2, we have
$$x^{2n}-1=\widetilde{f}_1(x)^2\widetilde{f}_2(x)^2\ldots \widetilde{f}_r(x)^2.$$
Since $f_1(x),f_2(x),\ldots,f_r(x)$ are pairwise coprime monic irreducible polynomials in $\mathbb{F}_{2^m}[x]$,
 $\widetilde{f}_1(x),\widetilde{f}_2(x),\ldots, \widetilde{f}_r(x)$  are pairwise coprime irreducible polynomials in $\mathbb{F}_{2^m}[x]$ as well. Hence for each integer $j$, $1\leq j\leq r$,
there is a unique integer $j^{\prime}$, $1\leq j^{\prime}\leq r$, such that $\widetilde{f}_j(x)=e_jf_{j^{\prime}}(x)$
where $e_j\in \mathbb{F}_{2^m}^{\times}$.
  We assume that $f_j(x)=\sum_{l=0}^{d_j}c_lx^l$ where $c_j\in \mathbb{F}_{2^m}$. Then
$x^{2d_j}f_j(x^{-1})^2=\widetilde{f}_j(x)^2$. From this, by Equation (4) and $x^{2n}=1$ in ${\cal A}$ we deduce
\begin{eqnarray*}
\varrho(\varepsilon_j(x))&=&1-x^{2n-({\rm deg}(b_j(x))+2d_j)}(x^{{\rm deg}(b_j(x))}v_j(x^{-1}))
(x^{2d_j}f_j(x^{-1})^2)\\
  &=&1-x^{2n-({\rm deg}(b_j(x))+2d_j)}\widetilde{b}_j(x)\widetilde{f}_j(x)^2\\
  &=&1-h_j(x)f_{j^{\prime}}(x)^2
\end{eqnarray*}
where $h_j(x)=e_j^2x^{2n-({\rm deg}(b_j(x))+2d_j)}\widetilde{b}_j(x)\in {\cal A}$.
Similarly, by (4) it
follows that $\varrho(\varepsilon_j(x))=g_j(x)F_{j^{\prime}}(x)$ for some $g_j(x)\in {\cal A}$. Then from these
and by Equations (3) and (4) we deduce that $\varrho(\varepsilon_j(x))=\varepsilon_{j^{\prime}}(x)$.

\par
   As stated above, we see that
for each integer $j$, $1\leq j\leq r$, there is a unique integer $j^{\prime}$, $1\leq j^{\prime}\leq r$, such that $\varrho(\varepsilon_j(x))=
\varepsilon_{j^{\prime}}(x)$. We still use $\varrho$ to denote this map $j\mapsto j^{\prime}$; i.e., $\varrho(\varepsilon_j(x))=\varepsilon_{\varrho(j)}(x)$.
Whether $\varrho$ denotes the ring automorphism of ${\cal A}$ or this map on the set $\{1,\ldots,r\}$ is determined by context.
The next lemma shows the compatibility of the two uses of $\varrho$.

\vskip 3mm \noindent
  {\bf Lemma 3.2} \textit{With the notations above, we have the following}:

\vskip 2mm \par
   (i) \textit{$\varrho$ is a permutation on $\{1,\ldots,r\}$ satisfying $\varrho^{-1}=\varrho$}.

\vskip 2mm \par
   (ii) \textit{After a rearrangement of $\varepsilon_1(x),\ldots,\varepsilon_r(x)$ there are integers $\lambda,\epsilon$ such that
$\varrho(j)=j$ for all $j=1,\ldots,\lambda$ and $\varrho(\lambda+l)=\lambda+\epsilon+l$ for all $l=1,\ldots,\epsilon$, where $\lambda\geq 1,
\epsilon\geq 0$ and $\lambda+2\epsilon=r$}.

\vskip 2mm\par
   (iii) \textit{For each integer $j$, $1\leq j\leq r$, there is a unique invertible element $e_j$ of
$\mathbb{F}_{2^m}$ such that $\widetilde{f}_j(x)=e_j f_{\varrho(j)}(x)$}.

\vskip 2mm \par
   (iv) \textit{For any integer $j$, $1\leq j\leq r$, $\varrho(\varepsilon_j(x))=\varepsilon_{\varrho(j)}(x)$ in the ring ${\cal A}$,
and $\varrho({\cal A}_{j})={\cal A}_{\varrho(j)}$. Then $\varrho$ induces a ring isomorphism from ${\cal A}_{j}[u]/\langle u^k\rangle$
 onto ${\cal A}_{\varrho(j)}[u]/\langle u^k\rangle$ in the natural way}.

\vskip 3mm \noindent
  \textit{Proof.} (i)--(iii) follow from the definition of the map $\varrho$, and
(iv) follows from that ${\cal A}_j=\varepsilon_j(x){\cal A}$ immediately.
\hfill $\Box$

\vskip 3mm \noindent
  {\bf Lemma 3.3} \textit{Let $\alpha(x)=\sum_{j=1}^r\alpha_j(x), \beta(x)=\sum_{j=1}^r\beta_j(x)\in {\cal A}[u]/\langle u^k\rangle$,
where $\alpha_j(x), \beta_j(x)\in{\cal A}_j[u]/\langle u^k\rangle$. Then}
$\alpha(x)\varrho(\beta(x))=\sum_{j=1}^r\alpha_j(x)\varrho(\beta_{\varrho(j)}(x)).$

\vskip 3mm \noindent
  \textit{Proof.} By Lemma 3.2 we have $\varrho(\beta_{\varrho(j)}(x))\in \varrho ({\cal A}_{\varrho(j)}[u]/\langle u^k\rangle)={\cal A}_{\varrho^2(j)}[u]/\langle u^k\rangle$ $={\cal A}_j[u]/\langle u^k\rangle$.
Hence $\alpha_j(x)\varrho(\beta_{\varrho(j)}(x))\in {\cal A}_j[u]/\langle u^k\rangle$ for all $j$. If
$l\neq \varrho(j)$, then $j\neq\varrho(l)$ and ${\cal A}_j{\cal A}_{\varrho(l)}=\{0\}$ by Lemma 2.1(ii), which implies $\alpha_j(x)\varrho(b_l(x))\in ({\cal A}_j[u]/\langle u^k\rangle)
({\cal A}_{\varrho(l)}[u]/\langle u^k\rangle)
=\{0\}$. Hence

\par
\ \ \  $\alpha(x)\varrho(\beta(x))=\sum_{j=1}^r\sum_{l=1}^r\alpha_j(x)\varrho(\beta_{l}(x))=\sum_{j=1}^r\alpha_j(x)\varrho(\beta_{\varrho(j)}(x))$.
\hfill $\Box$

\vskip 3mm \noindent
  {\bf Lemma 3.4} \textit{Using the notations above, for any $1\leq j\leq r$ we have}

\vskip 2mm \par
  (i) \textit{For any $\alpha=a(x)\in {\cal K}_j$, we denote $\widehat{\alpha}=(\varphi_{\varrho(j)}^{-1}\varrho\varphi_j)(\alpha)$. Then
$\widehat{ \ }$ is a ring isomorphism from ${\cal K}_j$ onto ${\cal K}_{\varrho(j)}$ such that the following diagram commutes}
$$\begin{array}{ccc} \ \ \ \ {\cal K}_j=\mathbb{F}_{2^m}[x]/\langle f_j(x)^2\rangle & \stackrel{\widehat{ \ }}{\longrightarrow} &  {\cal K}_{\varrho(j)}=\mathbb{F}_{2^m}[x]/\langle f_{\varrho(j)}(x)^2\rangle \cr
  \varphi_j  \downarrow &  & \ \ \ \downarrow \varphi_{\varrho(j)} \cr
 \ \ \ \ {\cal A}_j & \stackrel{\varrho}{\longrightarrow} &  {\cal A}_{\varrho(j)}
\end{array}$$
\textit{Specifically, we have $\widehat{\alpha}=a(x^{-1})\in {\cal K}_{\varrho(j)}$. Hence $\widehat{f}_j(x)=f_j(x^{-1})\in {\cal K}_{\varrho(j)}$}.

\vskip 2mm \par
  (ii) \textit{For any $\xi=\sum_{l=0}^{k-1}\alpha_lu^l\in {\cal K}_j[u]/\langle u^k\rangle$
where $\alpha_0,\alpha_1,\ldots,\alpha_{k-1}\in {\cal K}_j$, define $\widehat{\xi}=\sum_{l=0}^{k-1}\widehat{\alpha_l}u^l$. Then
$\varrho$ induces a ring isomorphism from ${\cal K}_j[u]/\langle u^{k}\rangle$ onto ${\cal K}_{\varrho(j)}[u]/\langle u^{k}\rangle$ by the rule that}
$\varrho: \xi\mapsto \widehat{\xi}$ $(\forall \xi\in {\cal K}_j[u]/\langle u^k\rangle)$.

\vskip 3mm \noindent
  \textit{Proof} (i) It follows from Lemma 2.1(iii) and Lemma 3.2 (iv).

\par
  (ii) It follows from (i) immediately.
\hfill $\Box$

\vskip 3mm \par
   Now, we give the dual code of each cyclic code over $R$ of length $2n$.

\vskip 3mm \noindent
   {\bf Theorem 3.5} \textit{Let ${\cal C}$ be a cyclic code
over $R$ of length $2n$ with canonical form decomposition
${\cal C}=\oplus_{j=1}^r\varepsilon_j(x)C_j$, where $C_j$ is an ideal of ${\cal K}_j[u]/\langle u^k\rangle$ given by Theorem 2.6. Then
the dual code ${\cal C}^{\bot_E}$ of ${\cal C}$ is given by}
$${\cal C}^{\bot_E}=\oplus_{j=1}^r\varepsilon_j(x)D_j,$$
\textit{where $D_j$ is an ideal of ${\cal K}_j[u]/\langle u^k\rangle$ given by one of the following eight cases for all
$j=1,\ldots,r$}:
{\small\begin{center}
\begin{tabular}{lll}\hline
case &  $C_j$  &  $D_{\varrho(j)}$  \\ \hline
1.   & $\langle u^i\rangle$ \ $(0\leq i\leq k)$ &  $\langle u^{k-i}\rangle$  \\
2.   &  $\langle u^sf_j(x)\rangle$ \ ($0\leq s\leq k-1$) & $\langle u^{k-s},f_{\varrho(j)}(x)\rangle$ \\
3.   & $\langle u^i+u^tf_j(x)\omega\rangle$ & $\langle u^{k-i}+u^{k+t-2i}f_{\varrho(j)}(x)\omega^{\prime}\rangle$ \\
     & ($\omega\in ({\cal F}_j[u]/\langle u^{i-t}\rangle)^{\times}$, $t\geq 2i-k$, & $\omega^{\prime}=e_jx^{2n-d_j}\widehat{\omega}$
       (mod $f_{\varrho(j)}(x)$)\\
     & $0\leq t<i\leq k-1$) & \\
4.   & $\langle u^i+f_j(x)\omega\rangle$  & $\langle u^i+f_{\varrho(j)}(x)\omega^{\prime}\rangle$ \\
     & $(\omega\in ({\cal F}_j[u]/\langle u^{k-i}\rangle)^{\times}$, $2i>k$ & $\omega^{\prime}=e_jx^{2n-d_j}\widehat{\omega}$
       (mod $f_{\varrho(j)}(x)$)\\
     & $0<i\leq k-1$) & \\
5.   & $\langle u^i+u^{t}f_j(x)\omega\rangle$ & $\langle u^{i-t}+f_{\varrho(j)}(x)\omega^{\prime},u^{k-i}f_{\varrho(j)}(x)\rangle$  \\
     & $(\omega\in ({\cal F}_j[u]/\langle u^{k-i}\rangle)^{\times}$, $t<2i-k$,   & $\omega^{\prime}=e_jx^{2n-d_j}\widehat{\omega}$
       (mod $f_{\varrho(j)}(x)$) \\
     & $1\leq t<i\leq k-1$)     & \\
6.   & $\langle u^{i},u^{s}f_j(x)\rangle$ \ $(0\leq s<i\leq k-1)$ & $\langle u^{k-s},u^{k-i}f_{\varrho(j)}(x)\rangle$ \\
7.   & $\langle u^{i}+f_j(x)\omega, u^{s}f_j(x)\rangle$ & $\langle u^{k-s}+u^{k-i-s}f_{\varrho(j)}(x)\omega^{\prime}\rangle$ \\
     & $(\omega\in ({\cal F}_j[u]/\langle u^s\rangle)^{\times}$, $i+s\leq k-1$,  & $\omega^{\prime}=e_jx^{2n-d_j}\widehat{\omega}$
       (mod $f_{\varrho(j)}(x)$) \\
     & $1\leq s<i\leq k-1$) & \\
8. & $\langle u^{i}+u^{t}f_j(x)\omega, u^{s}f_j(x)\rangle$ & $\langle u^{k-s}+u^{k+t-i-s}f_{\varrho(j)}(x)\omega^{\prime},u^{k-i}f_{\varrho(j)}(x)\rangle$ \\
   & $(\omega\in ({\cal F}_j[u]/\langle u^{s-t}\rangle)^{\times}$,  &  $\omega^{\prime}=e_jx^{2n-d_j}\widehat{\omega}$
      (mod $f_{\varrho(j)}(x)$) \\
   &  $i+s\leq k+t-1$,  & \\
   & $1\leq t<s<i\leq k-1)$ & \\ \hline
\end{tabular}
\end{center}}

\vskip 3mm \noindent
  \textit{Proof} Let $1\leq j\leq r$, and
$E_j$ be an ideal of ${\cal K}_j[u]/\langle u^k\rangle$ given by one of the following eight cases:

\par
  (i) $E_j=\langle u^{k-i}\rangle$,  if $C_j=\langle u^i\rangle$, where $0\leq i\leq k$.

\par
  (ii) $E_j=\langle u^{k-s},f_j(x)\rangle$,  if $C_j=\langle u^sf_j(x)\rangle$, where $0\leq s\leq k-1$.

\par
  (iii)  $E_j=\langle u^{k-i}-u^{k+t-2i}f_j(x)\omega\rangle$, if $C_j=\langle u^i+u^tf_j(x)\omega\rangle$ where $\omega\in ({\cal F}_j[u]/\langle u^{i-t}\rangle)^{\times}$,
 $0\leq t<i\leq k-1$ and $t\geq 2i-k$.

\par
  (iv) $E_j=\langle u^i-f_j(x)\omega\rangle$,  if $C_j=\langle u^i+f_j(x)\omega\rangle$ where $\omega\in ({\cal F}_j[u]/\langle u^{k-i}\rangle)^{\times}$,
$0<i\leq k-1$ and $2i>k$.

\par
  (v) $E_j=\langle u^{i-t}-f_j(x)\omega,u^{k-i}f_j(x)\rangle$,  if $C_i=\langle u^i+u^{t}f_j(x)\omega\rangle$ where
  $\omega\in ({\cal F}_j[u]/\langle u^{k-i}\rangle)^{\times}$,
$1\leq t<i\leq k-1$ and $t<2i-k$.

\par
  (vi) $E_j=\langle u^{k-s},u^{k-i}f_j(x)\rangle$, if $C_j=\langle u^{i},u^{s}f_j(x)\rangle$, where
$0\leq s<i\leq k-1$.

\par
 (vii) $E_j=\langle u^{k-s}-u^{k-i-s}f_j(x)\omega\rangle$,  if $C_j=\langle u^{i}+f_j(x)\omega, u^{s}f_j(x)\rangle$, where $\omega\in ({\cal F}_j[u]/\langle u^s\rangle)^{\times}$,
$1\leq s<i\leq k-1$, $i+s\leq k-1$.

\par
  (viii) $E_j=\langle u^{k-s}-u^{k+t-i-s}f_j(x)\omega,u^{k-i}f_j(x)\rangle$, when $C_j=\langle u^{i}+u^{t}f_j(x)\omega$, $u^{s}f_j(x)\rangle$, where $\omega\in ({\cal F}_j[u]/\langle u^{s-t}\rangle)^{\times}$,
$1\leq t<s<i\leq k-1$, $i+s\leq k+t-1$.

\vskip 2mm \noindent
  Then by a direct calculation we deduce that
$C_jE_j=\{a(x)b(x)\mid a(x)\in C_j, \ b(x)\in E_j\}=\{0\}$.
Further, by Theorem 2.6 we have $|C_j||E_j|=2^{2md_jk}$.

\par
   Since $\varrho$ induces a ring isomorphism from ${\cal K}_j[u]/\langle u^k\rangle$ onto ${\cal K}_{\varrho(j)}[u]/\langle u^k\rangle$ by
Lemma 3.4(ii), $\varrho(E_j)$ is an ideal of ${\cal K}_{\varrho(j)}[u]/\langle u^k\rangle$
and $|\varrho(E_j)|=|E_j|$. We
denote $D_{\varrho(j)}=\varrho(E_j)$. Let $l=\varrho(j)$. Then $j=\varrho(l)$ by $\varrho^2=\varrho$. Now, set
${\cal D}=\sum_{j=1}^r\varepsilon_{\varrho(j)}(x) D_{\varrho(j)}
=\oplus_{l=1}^r\varepsilon_l(x)D_{l}$, where $D_{l}=\varrho(E_{\varrho(l)})$.
By Theorem 2.3, we see that ${\cal D}$ is a cyclic code over $R$ of length $2n$.

\par
   For any $1\leq j\leq r$, by Lemma 2.1(iii) we see that $\varepsilon_{\varrho(j)}(x) D_{\varrho(j)}$ is an ideal of
${\cal A}_{\varrho(j)}[u]/\langle u^k\rangle$. Moreover, we have $\varrho(D_{\varrho(j)})=\varrho^2(E_j)=E_j$,
and $\varrho(\varepsilon_{\varrho(j)}(x))=\varepsilon_j(x)$ by Lemma 3.2(iv). From these
and by Lemma 3.3, we deduce that
\begin{eqnarray*}
{\cal C}\cdot \varrho({\cal D})&=&\left(\sum_{j=1}^r\varepsilon_j(x)C_j\right) \varrho\left(\sum_{l=1}^r\varepsilon_l(x)D_l\right)
    =\sum_{j=1}^r(\varepsilon_j(x)C_j) \varrho\left(\varepsilon_{\varrho(j)}(x)D_{\varrho(j)}\right)\\
    &=&\sum_{j=1}^r(\varepsilon_j(x)C_j) (\varepsilon_j(x)E_j)=\sum_{j=1}^r\varepsilon_j(x)(C_j E_j)=\{0\},
\end{eqnarray*}
which implies ${\cal D}\subseteq {\cal C}^{\bot_E}$ by Lemma 3.1.
  On the other hand, by Lemma 2.1(iii) and Theorem 2.3(iii), we have
$|{\cal C}||{\cal D}|=(\prod_{j=1}^r|C_j|)(\prod_{j=1}^r|E_j|)=\prod_{j=1}^r(|C_j||E_j|)
=\prod_{j=1}^r2^{2md_jk}=2^{2mk\sum_{j=1}^rd_j}=(2^{mk})^{2n}=|R[x]/\langle x^{2n}-1\rangle|$.
Since
$R=\mathbb{F}_{2^m}[u]/\langle u^k\rangle$ is a finite Frobenius ring, from the theory of linear codes over Frobenius rings (See [6],
for example)
we deduce that ${\cal C}^{\bot_E}={\cal D}$.

\par
  Finally, we give the generator set of $D_{\varrho(j)}=\varrho(E_j)$, $1\leq j\leq r$.
By Lemma 3.2(iii), Lemma 3.4 and $x^{2n}=1$, we have
$\varrho(u^l)=u^l$ for all $1\leq l\leq k$, $\varrho(\omega)=\widehat{\omega}$ for
any $\omega\in {\cal K}_j[u]/\langle u^k\rangle$ and $\varrho(f_j(x))=f_j(x^{-1})=x^{2n-d_j}(x^{d_j}f_j(x^{-1}))
=x^{2n-d_j}\widetilde{f}_j(x)=e_jx^{2n-d_j}f_{\varrho(j)}(x)$. Hence
$\varrho(u^lf_j(x)\omega)=\varrho(u^l)\varrho(f_j(x))\varrho(\omega)$ $=u^lf_{\varrho(j)}(x)\omega^{\prime}$,
where $\omega^{\prime}=e_jx^{2n-d_j}\widehat{\omega}\in {\cal F}_{\varrho(j)}[u]/\langle u^k\rangle$.
Then the conclusions follows from (i)--(viii) immediately.
\hfill $\Box$

\vskip 3mm \par
   Finally, using Theorems 2.3, 2.6 and 3.5 we list all distinct self-dual cyclic
codes over the ring $R$ of length $2n$ by the following Theorem.

\vskip 3mm
\noindent
  {\bf Theorem 3.6} \textit{Using the notations in Lemma 3.2$({\rm ii})$, all
distinct self-dual cyclic
codes over the ring $R$ of length $2n$ are give by}:
$${\cal C}=\left(\oplus_{j=1}^\lambda \varepsilon_j(x)C_j\right)\oplus
\left(\oplus_{j=\lambda+1}^{\lambda+\epsilon}(\varepsilon_{j}(x)C_{j}\oplus\varepsilon_{j+\epsilon}(x)C_{j+\epsilon})\right),$$
\textit{where $C_j$ is an ideal of ${\cal K}_j[u]/\langle u^k\rangle$ determined by the following conditions}:

\vskip 2mm\par
  (A) \textit{If $1\leq j\leq \lambda$, $C_j$ is determined by the following conditions}:

\vskip 2mm\par
  (A-i) \textit{When $k$ is even, $C_j$ is given by one of the following six cases}:

  \vskip 2mm \par
  (A-i-1) \textit{$C_j=\langle u^{\frac{k}{2}}\rangle$}.

\vskip 2mm \par
  (A-i-2) \textit{$C_j=\langle f_j(x)\rangle$}.

\vskip 2mm \par
  (A-i-3) \textit{$C_j=\langle u^{\frac{k}{2}}+u^{t}f_j(x)\omega\rangle$, where
$0\leq t\leq \frac{k}{2}-1$ and $\omega\in ({\cal F}_j[u]/\langle u^{\frac{k}{2}-t}\rangle)^{\times}$ satisfying
$\omega+e_jx^{2n-d_j}\widehat{\omega}\equiv 0$ $($mod $f_j(x)$$)$}.

\vskip 2mm \par
  (A-i-4) \textit{$C_j=\langle u^{i}+f_j(x)\omega\rangle$, where
$\frac{k}{2}+1\leq i\leq k-1$ and $\omega\in ({\cal F}_j[u]/\langle u^{k-i}\rangle)^{\times}$ satisfying
$\omega+e_jx^{2n-d_j}\widehat{\omega}\equiv 0$ $($mod $f_j(x)$$)$}.

\vskip 2mm \par
  (A-i-5) \textit{$C_j=\langle u^i,u^{k-i}f_j(x)\rangle$, where
$\frac{k}{2}+1<i\leq k-1$}.

\vskip 2mm \par
  (A-i-6) \textit{$C_j=\langle u^{i}+u^{t}f_j(x)\omega, u^{k-i}f_j(x)\rangle$, where
$1\leq t<k-i$, $\frac{k}{2}+1\leq i\leq k-1$ and
$\omega\in ({\cal F}_j[u]/\langle u^{k-i-t}\rangle)^{\times}$ satisfying
$\omega+e_jx^{2n-d_j}\widehat{\omega}\equiv 0$ $($mod $f_j(x)$$)$}.

\vskip 2mm\par
  (A-ii) \textit{When $k$ is odd, $C_j$ is given by one of the following four cases}:

\vskip 2mm \par
  (A-ii-1) \textit{$C_j=\langle f_j(x)\rangle$}.

\vskip 2mm \par
  (A-ii-2) \textit{$C_j=\langle u^{i}+f_j(x)\omega\rangle$, where
$\frac{k+1}{2}\leq i\leq k-1$ and $\omega\in ({\cal F}_j[u]/\langle u^{k-i}\rangle)^{\times}$ satisfying
$\omega+e_jx^{2n-d_j}\widehat{\omega}\equiv 0$ $($mod $f_j(x)$$)$}.

\vskip 2mm \par
  (A-ii-3) \textit{$C_j=\langle u^i,u^{k-i}f_j(x)\rangle$, where
$\frac{k+1}{2}<i\leq k-1$}.

\vskip 2mm \par
  (A-ii-4) \textit{$C_j=\langle u^{i}+u^{t}f_j(x)\omega, u^{k-i}f_j(x)\rangle$, where
$1\leq t<k-i$, $\frac{k+1}{2}\leq i\leq k-1$ and
$\omega\in ({\cal F}_j[u]/\langle u^{k-i-t}\rangle)^{\times}$ satisfying
$\omega+e_jx^{2n-d_j}\widehat{\omega}\equiv 0$ $($mod $f_j(x)$$)$}.

\vskip 2mm \par
  (B) \textit{If $j=\lambda+l$ where $1\leq l\leq \epsilon$, then $(C_j, C_{j+\epsilon})$ is given by
one of the following eight cases}:
{\small\begin{center}
\begin{tabular}{lll}\hline
case &  $C_j$  &  $C_{j+\epsilon}$  \\ \hline
B-1.   & $\langle u^i\rangle$ \ $(0\leq i\leq k)$ &  $\langle u^{k-i}\rangle$  \\
B-2.   &  $\langle u^sf_j(x)\rangle$ \ ($0\leq s\leq k-1$) & $\langle u^{k-s},f_{j+\epsilon}(x)\rangle$ \\
B-3.   & $\langle u^i+u^tf_j(x)\omega\rangle$ & $\langle u^{k-i}+u^{k+t-2i}f_{j+\epsilon}(x)\omega^{\prime}\rangle$ \\
     & ($\omega\in ({\cal F}_j[u]/\langle u^{i-t}\rangle)^{\times}$, & $\omega^{\prime}=e_jx^{2n-d_j}\widehat{\omega}$
       (mod $f_{j+\epsilon}(x)$)\\
     &  $t\geq 2i-k, 0\leq t<i\leq k-1$) & \\
B-4.   & $\langle u^i+f_j(x)\omega\rangle$  & $\langle u^i+f_{j+\epsilon}(x)\omega^{\prime}\rangle$ \\
     & $(\omega\in ({\cal F}_j[u]/\langle u^{k-i}\rangle)^{\times}$, & $\omega^{\prime}=e_jx^{2n-d_j}\widehat{\omega}$
       (mod $f_{j+\epsilon}(x)$)\\
     & $2i>k, 0<i\leq k-1$) & \\
B-5.   & $\langle u^i+u^{t}f_j(x)\omega\rangle$ & $\langle u^{i-t}+f_{j+\epsilon}(x)\omega^{\prime},u^{k-i}f_{j+\epsilon}(x)\rangle$  \\
     & $(\omega\in ({\cal F}_j[u]/\langle u^{k-i}\rangle)^{\times}$,    & $\omega^{\prime}=e_jx^{2n-d_j}\widehat{\omega}$
       (mod $f_{j+\epsilon}(x)$) \\
     & $t<2i-k, 1\leq t<i\leq k-1$)     & \\
B-6.   & $\langle u^{i},u^{s}f_j(x)\rangle$ & $\langle u^{k-s},u^{k-i}f_{j+\epsilon}(x)\rangle$ \\
       & $(0\leq s<i\leq k-1)$  & \\
B-7.   & $\langle u^{i}+f_j(x)\omega, u^{s}f_j(x)\rangle$ & $\langle u^{k-s}+u^{k-i-s}f_{j+\epsilon}(x)\omega^{\prime}\rangle$ \\
     & $(\omega\in ({\cal F}_j[u]/\langle u^s\rangle)^{\times}$,   & $\omega^{\prime}=e_jx^{2n-d_j}\widehat{\omega}$
       (mod $f_{j+\epsilon}(x)$) \\
     & $i+s\leq k-1, 1\leq s<i\leq k-1$) & \\
B-8. & $\langle u^{i}+u^{t}f_j(x)\omega, u^{s}f_j(x)\rangle$ & $\langle u^{k-s}+u^{k+t-i-s}f_{j+\epsilon}(x)\omega^{\prime},u^{k-i}f_{j+\epsilon}(x)\rangle$ \\
   & $(\omega\in ({\cal F}_j[u]/\langle u^{s-t}\rangle)^{\times}$,  &  $\omega^{\prime}=e_jx^{2n-d_j}\widehat{\omega}$
      (mod $f_{j+\epsilon}(x)$) \\
   &  $i+s\leq k+t-1$,  & \\
   & $1\leq t<s<i\leq k-1)$ & \\ \hline
\end{tabular}
\end{center}}


\section{An Example}
\noindent
    We consider cyclic codes of length $14$ over
$R=\mathbb{F}_2[u]/\langle u^4\rangle=\mathbb{F}_2+u\mathbb{F}_2+u^2\mathbb{F}_2+u^3\mathbb{F}_2$ ($u^4=0$)
which is a finite chain ring of $16$ elements.

\par
  In this case, we have $m=1$, $k=4$ and $n=7$. It is known that $x^7-1=f_1(x)f_2(x)f_3(x)$
where $f_1(x)=x+1$, $f_2(x)=x^3+x+1$ and $f_3(x)=x^3+x^2+1$ are irreducible polynomials in $\mathbb{F}_2[x]$
satisfying $\widetilde{f}_1(x)=f_1(x)$ and $\widetilde{f}_2(x)=f_3(x)$.
Hence $r=3$, $d_1=1$, $d_2=d_3=3$, $\lambda=1$ and $\epsilon=1$.

\par
   By Corollary 2.7, the number of cyclic codes over
$R$ of length $14$ is
\begin{center}
$\prod_{j=1}^3N_{(2,d_j,4)}=\prod_{j=1}^3\left(1+14+\Omega_1(2^{d_j},4)+\Omega_2(2^{d_j},4)+(2^{d_j}-1)\Gamma(2^{d_j},4)\right)$,
\end{center}
where $\Gamma(2^{d_j},4)=1$ for all $j=1,2,3$, $\Omega_1(2^{d_1},4)=5$, $\Omega_2(2^{d_1},4)=2$,
$\Omega_1(2^{d_j},4)=77$ and $\Omega_1(2^{d_j},4)=14$ for $j=2,3$. Then we have
$\prod_{j=1}^3N_{(2,d_j,4)}=23\cdot 113^2=293687$.
    Precisely, by Theorem 2.6 all distinct cyclic codes over
$R$ of length $14$ are given by
$${\cal C}=\varepsilon_1(x)C_1\oplus\varepsilon_2(x)C_2\oplus\varepsilon_3(x)C_3 \ {\rm with} \ |{\cal C}|=|C_1||C_2||C_3|,$$
where

\par
  $\diamondsuit$ $\varepsilon_1(x)=x^{12}+x^{10}+x^{8}+x^6+x^4+x^2+1$, $\varepsilon_2(x)=x^8+x^4+x^2+1$ and $\varepsilon_3(x)=x^{12}+x^{10}
+x^6+1$.

\par
  $\diamondsuit$ $C_1$ is an ideal of the ring ${\cal K}_1[u]/\langle u^4\rangle$, where
${\cal K}_1=\mathbb{F}_2[x]/\langle (x+1)^2\rangle$ and ${\cal F}_1=\mathbb{F}_2[x]/\langle x+1\rangle=\mathbb{F}_2$,
given by the following table:
{\small\begin{center}
\begin{tabular}{llll}\hline
case &  number  &  $C_1$     &   $|C_1|$ \\ \hline
I.   & $5$  &  $\langle u^i\rangle$ \ $(0\leq i\leq 4)$ & $4^{4-i}$ \\
II.   & $4$    &   $\langle u^s (x+1)\rangle$ \ $(0\leq s\leq 3)$ &  $2^{4-s}$  \\
III.   & $5$ &    $\langle u+(x+1)\rangle$ &  $4^{3}$ \\
       &     &    $\langle u^2+(x+1)\rangle$, $\langle u^2+(x+1)(1+u)\rangle$, $\langle u^2+u(x+1)\rangle$ &  $4^{2}$ \\
      &      &  $\langle u^3+u^2(x+1)\rangle$ &  4 \\
IV.  & $2$     &    $\langle u^3+(x+1)\rangle$ &  $2^{4}$ \\
     &        &     $\langle u^3+u(x+1)\rangle$ &  $2^{3}$ \\
V.   &  $6$    &     $\langle u^i,u^s(x+1)\rangle$ $(0\leq s<i\leq 3)$ &  $2^{(8-(i+s))}$ \\
VI.   &  $1$    &     $\langle u^2+(x+1), u(x+1)\rangle$ &  $2^{5}$ \\
 \hline
\end{tabular}
\end{center}}

\par
  $\diamondsuit$ For $j=2,3$, $C_j$ is an ideal of ${\cal K}_j[u]/\langle u^4\rangle$, where
${\cal K}_j=\mathbb{F}_2[x]/\langle f_j(x)^2\rangle$ and ${\cal F}_j=\mathbb{F}_2[x]/\langle f_j(x)\rangle=\{a_0+a_1x+a_2x^2\mid
a_0,a_1,a_2\in\mathbb{F}_2\}$,
given by the following table:
{\small\begin{center}
\begin{tabular}{llll}\hline
case &  number  &  $C_j$     &   $|C_j|$ \\ \hline
I.   & $5$  & $\bullet$ $\langle u^i\rangle$ \ $(0\leq i\leq 4)$ & $64^{4-i}$ \\
II.   & $4$    & $\bullet$  $\langle u^s f_j(x)\rangle$ \ $(0\leq s\leq 3)$ &  $8^{4-s}$  \\
III.   & $7$ & $\bullet$   $\langle u+f_j(x)\omega\rangle$ ($\omega=a_0+a_1x+a_2x^2$, &  $64^{3}$ \\
       &     &   $ \ \   (a_0,a_1,a_2)\neq(0,0,0),\ a_0,a_1,a_2\in \mathbb{F}_2)$  & \\
       & $56$   & $\bullet$    $\langle u^2+f_j(x)\omega\rangle$ &  $64^{2}$ \\
       &     &   $ \ \  (\omega=a_0+a_1x+a_2x^2+u(b_0+b_1x+b_2x^2), $  & \\
       &     &   $  \ \ (a_0,a_1,a_2)\neq(0,0,0), \ a_0,a_1,a_2, b_0,b_1,b_2\in \mathbb{F}_2)$ & \\
       &  $7$   &  $\bullet$  $\langle u^2+uf_j(x)\omega\rangle$ ($\omega=a_0+a_1x+a_2x^2$, &  $64^{2}$ \\
       &     &   $ \ \  (a_0,a_1,a_2)\neq(0,0,0), \ a_0,a_1,a_2\in \mathbb{F}_2)$  & \\
      &  $7$ & $\bullet$  $\langle u^3+u^2f_j(x)\omega\rangle$ ($\omega=a_0+a_1x+a_2x^2$, &  64 \\
       &     &   $ \ \  (a_0,a_1,a_2)\neq(0,0,0), \ a_0, a_1,a_2\in \mathbb{F}_2)$  & \\
IV.  & $7$     &  $\bullet$  $\langle u^3+f_j(x)\omega\rangle$ ($\omega=a_0+a_1x+a_2x^2$, &  $8^{4}$ \\
    &     &   $ \ \  (a_0,a_1,a_2)\neq(0,0,0), \ a_0, a_1,a_2\in \mathbb{F}_2)$  & \\
     & $7$     &  $\bullet$   $\langle u^3+uf_j(x)\omega\rangle$ ($\omega=a_0+a_1x+a_2x^2$, &  $8^{3}$ \\
    &     &   $ \ \  (a_0,a_1,a_2)\neq(0,0,0), \ a_0, a_1,a_2\in \mathbb{F}_2)$  & \\
V.   &  $6$    &  $\bullet$   $\langle u^i,u^sf_j(x)\rangle$ $(0\leq s<i\leq 3)$ &  $8^{(8-(i+s))}$ \\
VI.   &  $7$    &  $\bullet$    $\langle u^2+f_j(x)\omega, uf_j(x)\rangle$ ($\omega=a_0+a_1x+a_2x^2$,&  $8^{5}$ \\
    &     &   $ \ \  (a_0,a_1,a_2)\neq(0,0,0), \ a_0, a_1,a_2\in \mathbb{F}_2)$  & \\
 \hline
\end{tabular}
\end{center}}

\vskip 3mm\par
   Moreover, by Theorem 3.6 all distinct self-dual cyclic codes over $R$ of length $14$ are given by
${\cal C}=\varepsilon_1(x)C_1\oplus\varepsilon_2(x)C_2\oplus\varepsilon_3(x)C_3$,
where

\par
  $\bullet$ $C_1$ is given by one of the following cases:

\par
  (A-i-1) $C_1=\langle u^{2}\rangle$.

\par
  (A-i-2) $C_1=\langle (x+1)\rangle$.

\par
  (A-i-3) $C_1=\langle u^2+(x+1)\rangle$, $C_1=\langle u^2+(x+1)(1+u)\rangle$; $C_1=\langle u^2+u(x+1)\rangle$.

\par
  (A-i-4) $C_1=\langle u^3+(x+1)\rangle$.

 \par
  (A-i-5) $C_1=\langle u^3,u(x+1)\rangle$.

\par
  $\bullet$ $(C_2,C_3)$ is given by one pair of ideals in the following table:
{\small\begin{center}
\begin{tabular}{lll}\hline
case &  $C_2$  &  $C_{3}$  \\ \hline
B-1.   & $\langle u^i\rangle$,  $0\leq i\leq 4$ &  $\langle u^{k-i}\rangle$  \\
B-2.   &  $\langle u^sf_2(x)\rangle$,  $0\leq s\leq 3$ & $\langle u^{4-s},f_{3}(x)\rangle$ \\
B-3.   &  $\langle u+f_2(x)\omega_1\rangle$, $\omega_1\in {\cal F}_2^{\times}$ & $\langle u^3+u^2f_3(x)\omega_1^{\prime}\rangle$,
         $\omega_1^{\prime}=x^{11}\widehat{\omega}_1$    \\
       &  $\langle u^2+f_2(x)\omega_2\rangle$, $\omega_2\in ({\cal F}_2[u]/\langle u^2\rangle)^{\times}$ &
          $\langle u^2+f_3(x)\omega_2^{\prime}\rangle$,
         $\omega_2^{\prime}=x^{11}\widehat{\omega}_2$   \\
       &  $\langle u^2+uf_2(x)\omega_1\rangle$, $\omega_1\in {\cal F}_2^{\times}$ & $\langle u^2+uf_3(x)\omega_1^{\prime}\rangle$,
         $\omega_1^{\prime}=x^{11}\widehat{\omega}_1$    \\
       &  $\langle u^3+u^2f_2(x)\omega_1\rangle$, $\omega_1\in {\cal F}_2^{\times}$ & $\langle u+f_3(x)\omega_1^{\prime}\rangle$,
         $\omega_1^{\prime}=x^{11}\widehat{\omega}_1$    \\
B-4.   &  $\langle u^3+f_2(x)\omega_1\rangle$, $\omega_1\in {\cal F}_2^{\times}$ & $\langle u^3+f_3(x)\omega_1^{\prime}\rangle$,
         $\omega_1^{\prime}=x^{11}\widehat{\omega}_1$    \\
B-5.   &  $\langle u^3+uf_2(x)\omega_1\rangle$, $\omega_1\in {\cal F}_2^{\times}$ & $\langle u^2+f_3(x)\omega_1^{\prime},uf_3(x)\rangle$,
         $\omega_1^{\prime}=x^{11}\widehat{\omega}_1$    \\
B-6.   & $\langle u^{i},u^{s}f_2(x)\rangle$, $0\leq s<i\leq 3$ & $\langle u^{4-s},u^{4-i}f_{3}(x)\rangle$ \\
B-7.   & $\langle u^2+f_2(x)\omega_1, uf_2(x)\rangle$, $\omega_1\in {\cal F}_2^{\times}$ & $\langle u^3+uf_3(x)\omega_1^{\prime}\rangle$,
         $\omega_1^{\prime}=x^{11}\widehat{\omega}_1$    \\
 \hline
\end{tabular}
\end{center}}

\noindent
in which $\omega_1=a_0+a_1x+a_2x^2$, $(a_0,a_1,a_2)\neq(0,0,0)$, $a_0,a_1,a_2\in \mathbb{F}_2$,
$\omega_2=a_0+a_1x+a_2x^2+u(b_0+b_1x+b_2x^2)$, $(a_0,a_1,a_2)\neq(0,0,0)$, $a_0,a_1,a_2,b_0,b_1,b_2\in \mathbb{F}_2$,
$\omega_1^{\prime}=x^{11}(a_0+a_1x^{-1}+a_2x^{-2})$ (mod $f_3(x)$) and
$\omega_2^{\prime}=x^{11}(a_0+a_1x^{-1}+a_2x^{-2}+u(b_0+b_1x^{-1}+b_2x^{-2}))$ (mod $f_3(x)$).

\par
  Therefore, the number of self-dual cyclic codes of length $14$ over $\mathbb{F}_2+u\mathbb{F}_2+u^2\mathbb{F}_2+u^3\mathbb{F}_2$
($u^4=0$) is equal to $7\cdot 113=791$.


\section{Conclusions and further research}
\noindent
We have developed a theory for cyclic codes of length $2n$ over the finite chain ring $R=\mathbb{F}_{2^m}[u]/\langle u^k\rangle=\mathbb{F}_{2^m}
+u\mathbb{F}_{2^m}+\ldots+u^{k-1}\mathbb{F}_{2^m}$ ($u^k=0$) for any integer $k\geq 2$ and positive odd integer $n$,
including the enumeration and construction of
these codes, the dual code and self-duality for each of these codes. These codes enjoy a rich algebraic structure compared
to arbitrary linear codes (which makes the search process much simpler).
Obtaining some bounds for minimal distance such as BCH-like of a cyclic code over the ring $R$ by just looking at the representation of such codes are future topics of interest.

\begin{acknowledgements}
  Part of this work was done when Yonglin Cao was visiting Chern Institute of Mathematics, Nankai University, Tianjin, China. Yonglin Cao would like to thank the institution for the kind hospitality. This research is
supported in part by the National Key Basic Research Program of China (Grant No. 2013CB834204) and the National Natural Science Foundation of
China (Grant No. 11471255).
\end{acknowledgements}



\end{document}